\documentclass[pdflatex,sn-mathphys-num]{sn-jnl}
\usepackage{natbib}
\usepackage{graphicx}%
\usepackage{multirow}%
\usepackage{amsmath,amssymb,amsfonts}%
\usepackage{amsthm}%
\usepackage{mathrsfs}%
\usepackage[title]{appendix}%
\usepackage{xcolor}%
\usepackage{textcomp}%
\usepackage{manyfoot}%
\usepackage{booktabs}%
\usepackage{algorithm}%
\usepackage{algorithmicx}%
\usepackage{algpseudocode}%
\usepackage{listings}%

\def\apj{Astrophys. J.}
\def\apjs{Astrophys. Suppl.}
\def\aj{Astron. J.}
\def\apjl{Astrophys. J. Lett.}
\def\apss{Astrophys. Space Sci.}

\def\nphysa{Nucl. Phys. A}
\def\pasp{Publ. Astron. Soc. Aust.}

\def\mnras{Mon. Not. R. Astron. Soc.}
\def\aap{Astron. Astrophys.}
\def\HII{{{H}{II}}}
\def\ratioR23{([OII]~$\lambda$3726 +[OIII]~$\lambda\lambda$4959,5007)/H$\beta$}
\def\R23{${\rm R}_{23}$}
\def\Msun{${\rm M}_{\odot}$}
\def\NII{[NII]}
\def\OH{$\log({\rm O/H})+12$}
\def\NIISII{[NII]/[SII]}
\def\SII{[SII]}
\def\Hb{H$\beta$}
\def\Ha{H$\alpha$}

\begin{document}

\title[Extragalactic Archaeology]{The assembly history of NGC 1365 through chemical archaeology}



\author*[1,2]{\fnm{Lisa} J. \sur{Kewley}}\email{lisa.kewley@cfa.harvard.edu}

\author*[2,3]{\fnm{Kathryn} \sur{Grasha}}\email{kathryn.grasha@anu.edu.au}

\author[4,5]{\fnm{Alex} \sur{Garcia}}

\author[5,6,7]{\fnm{Paul} \sur{Torrey}}

\author[8]{\fnm{Jeff} \sur{Rich}}

\author[9]{\fnm{Z. S.} \sur{Hemler}}

\author[2,3]{\fnm{Qian-Hui} \sur{Chen}}

\author[1,2]{\fnm{Peixin} \sur{Zhu}}

\author[8]{\fnm{Mark} \sur{Seibert}}

\author[1]{\fnm{Lars} \sur{Hernquist}}

\author[8]{\fnm{Barry} \sur{Madore}}

\affil*[1]{\orgdiv{Institute for Theory \& Computation}, \orgname{Center for Astrophysics | Harvard \& Smithsonian}, \orgaddress{\street{60 Garden Street}, \city{Cambridge}, \postcode{02138}, \state{MA}, \country{USA}}}

\affil*[2]{\orgdiv{Research School for Astronomy \& Astrophysics}, \orgname{Australian National University}, \orgaddress{\street{Cotter Road}, \city{Weston Creek}, \postcode{2611}, \state{ACT}, \country{Australia}}}

\affil[3]{\orgdiv{ARC Centre of Excellence for All-Sky Astrophysics in 3 Dimensions (ASTRO 3D)}, \orgname{Australian National University}, \orgaddress{\street{Cotter Road}, \city{Weston Creek}, \postcode{2611}, \state{ACT}, \country{Australia}}}

\affil[4]{\orgdiv{Department of Astronomy}, \orgname{University of Florida}, \orgaddress{\street{211 Bryant Space Sciences Center}, \city{Gainesville}, \postcode{32611}, \state{FL}, \country{USA}}}

\affil[5]{\orgdiv{Department of Astronomy}, \orgname{University of Virginia}, \orgaddress{\street{530 McCormick Road}, \city{Charlottesville}, \postcode{22904}, \state{VA}, \country{USA}}}

\affil[6]{\orgdiv{Virginia Institute for Theoretical Astronomy}, \orgname{University of Virginia}, \orgaddress{\street{530 McCormick Road}, \city{Charlottesville}, \postcode{22904}, \state{VA}, \country{USA}}}

\affil[7]{\orgdiv{The NSF-Simons AI Institute for Cosmic Origins}, \orgname{University of Virginia}, \orgaddress{\street{530 McCormick Road}, \city{Charlottesville}, \postcode{22904}, \state{VA}, \country{USA}}}

\affil[8]{\orgdiv{The Observatories}, \orgname{Carnegie Institution for Science}, \orgaddress{\street{813 Santa Barbara Street}, \city{Pasadena}, \postcode{91106}, \state{CA}, \country{USA}}}

\affil[9]{\orgdiv{Department of Astrophysical Sciences}, \orgname{Princeton University}, \orgaddress{\street{Peyton Hall}, \city{Princeton}, \postcode{08544}, \state{NJ}, \country{USA}}}


\abstract{Galaxies build through infalling gas and galaxy mergers. Tracking the dynamical history of a galaxy from a single snapshot in time is notoriously difficult.  We show that the dynamical history of a galaxy can be tracked using oxygen abundances as archeological tracers.  We derive the gas-phase oxygen abundances for 4546 spaxels across the face-on spiral galaxy NGC~1365 at a spatial resolution of 175 pc, providing one of the most detailed chemical fossil records of a spiral galaxy outside our Milky Way. We apply IllustrisTNG cosmological simulations to analyse the chemical abundance distribution in a theoretical model for NGC~1365. In the model, the main disk oxygen abundance gradient formed earliest, 11.9-12.5 billion years ago via mergers with multiple dwarf galaxies. A steep inner bar gradient formed slowly over the last 12 billion years through enrichment from star formation triggered by the infall of gas into the nuclear regions. An extended ionized gas disk with flat oxygen abundances was assembled more recently (5.9-8.6 billion years ago) through a minor merger.  This work indicates that cosmological simulations and ultra-high spatial resolution oxygen abundances together can provide an archaeological probe of the star formation and merger histories of spiral galaxies.}



\maketitle

\section{Main Text}\label{Intro}

\subsection{Introduction}

This study explores the complex metallicity gradient of the spiral galaxy NGC 1365 using a new ultra-high resolution technique for measuring spectra across galaxies in 3D. This method allows for a detailed examination of the galaxy's oxygen abundance structure at resolutions more than ten times finer than previous surveys. The research identifies three distinct oxygen abundance gradients across the galaxy. Historically, modeling galaxy oxygen abundance gradients has been limited by resolution constraints, but recent advances in simulations now enable a more detailed analysis across galaxies and through cosmic time, aligning closely with observational capabilities. This study utilizes the highest resolution volumes from the Next Generation Illustris theoretical simulations to investigate the origins of the multiple oxygen abundance components, offering insights into the possible star formation and gas accretion history of NGC 1365. 

Theoretical models predict that star formation in spiral galaxies occurs inside-out, with stars forming first in the central regions of the galaxy, and subsequently forming in the outer regions over the last 4-6 Gyr \cite{Fall80, Chiappini01, Somerville08}.  In this scenario, spiral galaxies initially form steep metallicity gradients with chemically enriched gas in their central regions and less enriched gas in their outer regions \cite{Kobayashi11, Gibson13, Molla19, Hemler21}.  As star formation progresses to the outer regions at later times, the gradients become more shallow \cite{Garcia23}.  This picture of inside-out star formation is consistent with the star-formation history and steep metallicity gradients seen in the Milky Way and nearby spiral galaxies \cite{Sanchez13, Stanghellini15, Franchetto21}.

The gas-phase oxygen abundance (referred to as the metallicity) distribution within a galaxy is modified over time by large-scale gas flows.  Galaxies accrete gas from the circumgalactic and intergalactic medium \cite{Perez11, Cameron21}, while galactic-scale metal-enriched outflows can selectively remove metals from galaxies \cite{Cameron21}.  Gas flows triggered by mergers flatten metallicity gradients \cite{Kewley06b, Rupke10, Perez11, Torrey12, Rosa14}, while minor mergers may introduce less enriched gas into a galaxy \cite{Zinchenko15}.  Tidal effects from cluster environments also modify metallicity gradients \cite{Franchetto21}.  

In the past, metallicity gradients in galaxies were measured using individual star-forming regions. Recently, integral field unit (IFU) data has facilitated the analysis of metallicity gradients using individual spaxels across galaxies.  However, the spatial resolution of most large IFU surveys is limited to 800~pc to 1~kpc, significantly larger than the size of individual star-forming regions ($\sim$50--100~pc).  In such data, an individual spaxel may contain the emission from several star-forming regions, as well as contamination from surrounding diffuse ionized gas.  This contamination can impact metallicity measurements and flatten metallicity gradients that are intrinsically steep \cite{Poetrodjojo19}.  
 
In this paper, we present spaxel-based metallicity analysis of NGC~1365 using 3D data obtained by the TYPHOON Survey using the Progressive Integral Step Method (PrISM) \cite{Sturch12}. PrISM provides a resolution of 175 pc or less, almost an order of magnitude improvement over previous IFU surveys, while the large effective field of view (18 arcmin long slit with 1.65 arcsec width stepped across an entire galaxy) allows us to observe the entire disc of nearby spiral galaxies in an IFU-like manner at an unprecedented spatial resolution of $\sim$4–5~pc in the closest galaxies, and an average spatial resolution between 50–100~pc.  

NGC~1365 is a spiral galaxy in the Fornax Cluster, containing a Seyfert 1.8 nucleus \cite{Veron-Cetty06}. The central star formation region of NGC~1365 is asymmetric with two large dust lanes \cite{Beck05} within an extended bar of 9.5 kpc radius \cite{Zurita21}.   The close distance of NGC~1365 \cite[18.1~Mpc][]{Jang18}, large size \cite[67.44 kpc diameter;][]{Lauberts82}, and mild inclination of $35.7^{\circ}$ \cite{Ho17} allows detailed analysis of the metallicity structure of NGC~1365 via the PrISM method. 

The metallicity gradient of NGC~1365 has been previously measured by several authors, with most authors fitting a single linear gradient \cite{Pagel79, Alloin81, Bresolin05, Dors05,Ho17}.  Roy et al. found a steep metallicity gradient of -0.05~dex~kpc$^{-1}$ inside a break radius of R=16.9~kpc, and flat metallicity distribution outside this radius \cite{Roy97}.  
The metallicity gradient in NGC 1365 was recently re-analysed in the context of comparisons between the metallicity gradients in barred and unbarred galaxies. Zurita et al. (2021) fit a single metallicity profile and find a metallicity gradient between $-0.004 \pm 0.001$~dex~kpc$^{-1}$ and $-0.014 \pm 0.002$~dex~kpc$^{-1}$, depending on the diagnostic \cite{Zurita21}.  Chen et al. (2023) adopt a two-component fit to the metallicity profile and find $-0.055$~dex~kpc$^{-1}$ inside the bar region, significantly steeper than the gradient in the disc region ($-0.0123$~dex~kpc$^{-1}$) with a change in the gradient occurring at 5.8~kpc\cite{QianhuiChen23}.  

 We measure the metallicity gradient in NGC~1365 using 4546 spaxels, providing $\sim$30 times the metallicity data used in previous gradient studies.  This data has sufficient radial resolution across 28 kpc in radius to resolve three distinct metallicity gradient sections.  Understanding such complex metallicity gradients has suffered from a lack of theoretical modelling at sufficient resolution for comparison with observations \cite[see][for a discussion of this issue in outer galaxy disks]{Bresolin17}.  Over the past few years, simulations have advanced significantly and can now model the metallicity gradients in galaxies across cosmic time at spatial resolutions consistent with the very best nearby 3D imaging spectroscopy \cite[e.g.,][]{Hemler21, Garcia23}. 

In this work, we use the highest resolution volume of The Next Generation (TNG) Illustris simulations \cite{Springel18} to explore the cause of the three metallicity components to NGC~1365.  This technique offers great promise for tracing the star formation and accretion history of nearby spiral galaxies.  

\subsection{Results}

\begin{figure*}[ht!]
\includegraphics[width=0.5\textwidth]{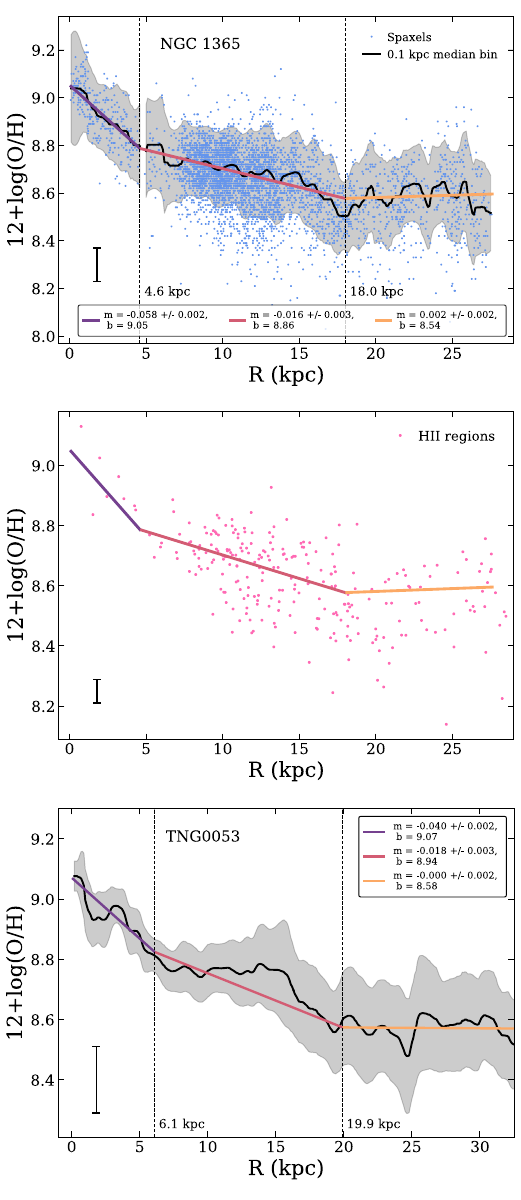}\\
\caption{{The metallicity as a function of radius in kiloparsecs for NGC~1365 and TNG0053.}
The metallicity gradient for 4546 spaxels (top panel) and 283 HII regions in NGC~1365 (middle panel) are shown in comparison with the metallicity gradient for the matched Illustris simulated galaxy TNG0053 (bottom panel).  Solid black lines show the median metallicity as a function of radius, and the shaded regions show the standard deviation from the median.  The three colored lines show the best-fit linear fits to the radial metallicity gradient. The vertical dashed lines show the location of the breaks of the linear fits. The radii have been corrected for inclination.   A representative error bar is shown in the lower left corner of each panel.  For NGC~1365, these error bars include the systematic uncertainties in the flux calibration, metallicity calibration and reddening correction. 
} 
\label{Fig_Gradient}
\end{figure*}

The method used to calculate the metallicity gradient in NGC~1365 is given in Section~\ref{Methods}.  \autoref{Fig_Gradient} shows the resulting radial metallicity distribution for the spaxels and HII regions in NGC~1365.  The metallicity distribution has three distinct zones: (1) an inner steep gradient corresponding to the bar region ($R\lesssim7$~kpc), (2) a main disk gradient corresponding to the star-forming regions along and between the spiral arms ($7<R<17$~kpc), and (3) a flat outer disk component that extends from $17$~kpc to 28~kpc.   We measure a gradient of $-0.058 \pm 0.002$~kpc$^{-1}$ in the steep inner bar region, $-0.016 \pm 0.003$~dex~kpc$^{-1}$ within the main disk, and $0.002 \pm 0.002$~dex~kpc$^{-1}$ in the flat outer disk.

Some studies of barred galaxies suggest that the presence of bars flattens metallicity gradients \cite{Zurita21}, other studies show a correlation between the presence of a bar and a steep gradient, or no correlation \cite{Zinchenko15}.  In NGC~1365, it is clear that a steep inner gradient can exist within the bar region.  This steep gradient is not caused by AGN contamination because (1) we have removed AGN-contaminated spaxels, and (2) we see the same steep inner gradient in the fit to the HII regions in NGC~1365 (\autoref{Fig_Gradient}, middle panel).

Our results support the metallicity gradient break radius delineating the inner bar region identified by Roy et al. (1997) \cite{Roy97} and the bar and main disk gradients measured by Chen et al. (2023) \cite{QianhuiChen23}.  We also identify a flat outer gradient, consistent with the flat outer gradients seen in the Milky Way and some nearby spiral galaxies \cite{Bresolin09, Sanchez-Menguiano18, Grasha22}.  

To understand the cause of the three-zone metallicity gradient in NGC 1365, we use the Illustris TNG cosmological simulation suite \cite{Springel18,Marinacci18}.
The TNG simulations include key astrophysical processes that impact gas-phase metallicity gradients, including star formation, chemical enrichment, radiative gas cooling, stellar, and supermassive black hole feedback.   We search the highest resolution TNG model suite (TNG50 with box size of 51.7 Mpc$^3$ containing $\sim 20,000$ resolved galaxies) for the closest theoretical model match to the metallicity gradient and stellar mass of NGC~1365 (see Section~\ref{Methods} for a description of the TNG models and match procedure).  The closest match, TNG0053, has a stellar mass of $\log({\rm M}_{*}/$\Msun$) = 10.67$ and a metallicity distribution (\autoref{Fig_Gradient}) that is well approximated by three zones, including an inner steep bar gradient, a shallow central region, and an outer flat metallicity gradient (reduced $\chi^2=0.123$).  A description of other close match simulated galaxies is given in Section~\ref{Methods}.

\begin{figure*}[t!]
\includegraphics[width=\textwidth]{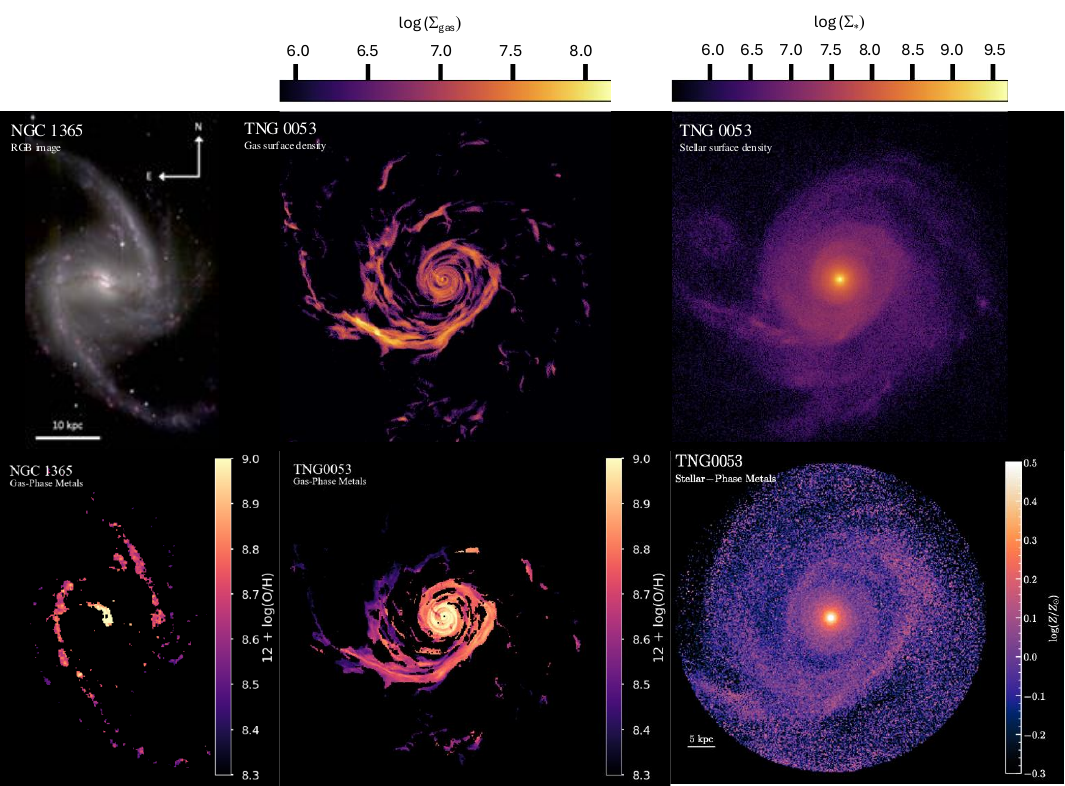}
\caption{{Comparison between NGC~1365 and TNG0053.}
The TYPHOON RGB image and gas-phase metallicity maps of NGC~1365 are compared with the face-on projections of the z=0 distributions in TNG0053 of gas and stellar surface density, gas-phase metallicity, and stellar metallicity.}
\label{fig:NGC_TNG0053_images}
\end{figure*}

\autoref{fig:NGC_TNG0053_images} shows the TYPHOON RGB image of NGC~1365 and the z=0 images of TNG0053 in gas surface density, stellar surface density, gas-phase metallicity, and stellar metallicity.  Despite the fact that NGC~1365 and TNG0053 were matched independently of their morphology, they both display grand-design spiral morphology with radial spokes in the spiral arm structure.  

Metallicity gradient breaks can be produced by a variety of mechanisms in theoretical models.  Previous theoretical simulations have produced gradient breaks via young bars \cite{Friedli95, Stanghellini15, Martel18}, corotation resonance in spiral arms \cite{Mishurov02}, minor and major mergers \cite{Zinchenko15,Buck23}, and feedback-generated outflows and mixing \cite{Wuyts16}.   Thus, the variety of metallicity gradients in NGC~1365 might arise from a combination of these effects. 

\begin{figure*}[t!]
\includegraphics[width=\textwidth]{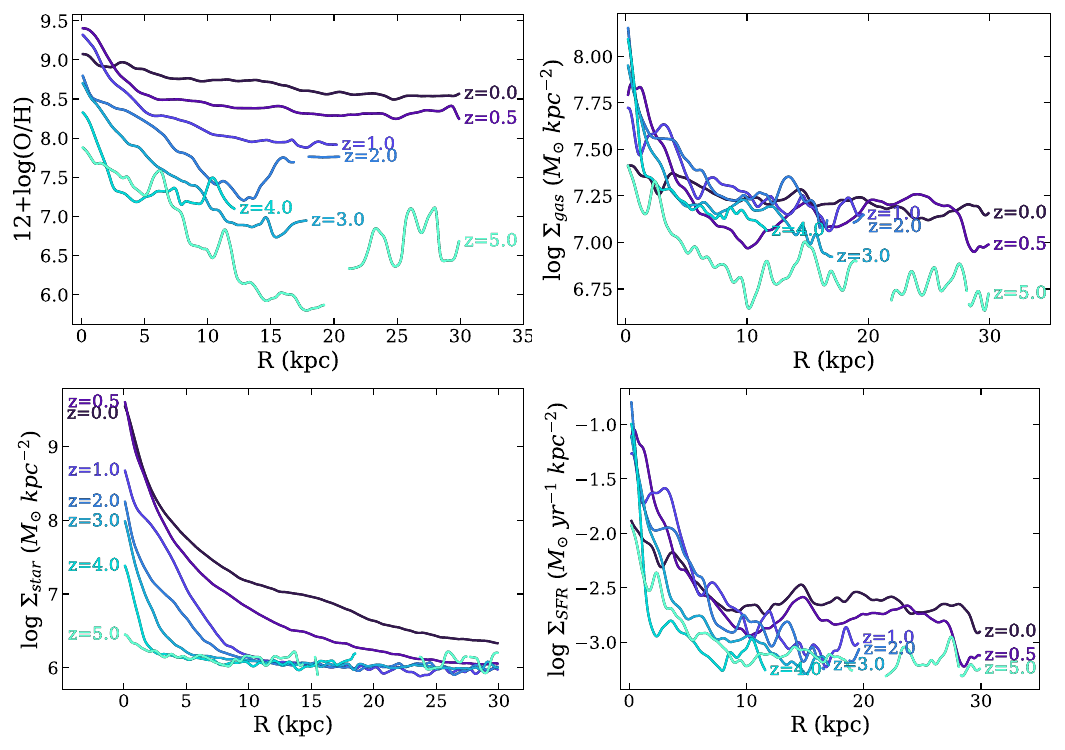}
\caption{{The metallicity gradient (top left) and gas (top right), stellar (bottom left), and star formation rate surface density (bottom right) of TNG0053.}
Radial distributions of the TNG0053 metallicity gradient, and the TNG0053 gas, stellar and star formation rate surface density from $z=5$ to $z=0$ are shown.}
\label{fig:Fig_Gradient_sims_z}
\end{figure*}

\begin{figure*}[t!]
\includegraphics[width=\textwidth]{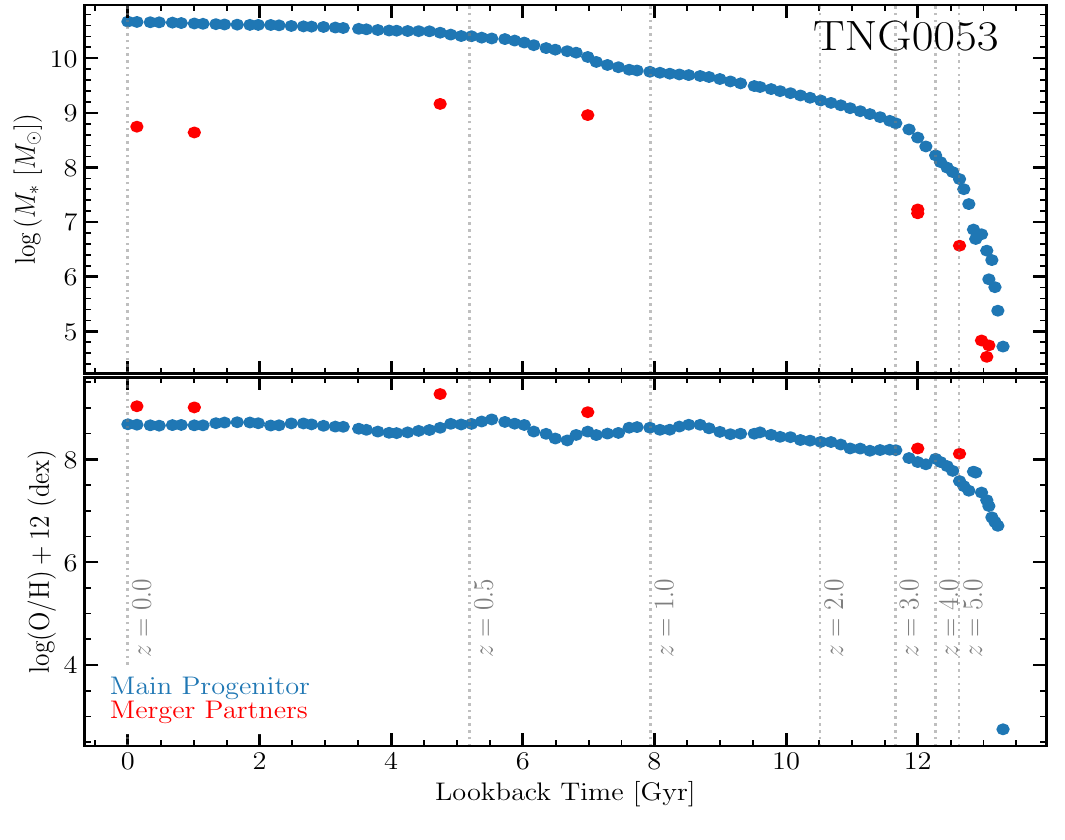}
\caption{{The stellar, metallicity, and merger history of TNG0053.}
Top: Stellar mass history of TNG0053 as a function of lookback time in Gyr.  Bottom: Metallicity history of TNG0053 as a function of lookback time in Gyr.  At early times ($z>3$) not every infalling galaxy in TNG50 has a metallicity value. In these cases, the infalling galaxies are metal poor such that no resolution element within the galaxy has a metallicity associated with it.  In both panels, the vertical dotted lines indicate redshifts of interest. The stellar masses of the central TNG0053 galaxy and large (relative to the central; mass fraction $>1/100$) galaxies merging with the primary galaxy are shown.} 
\label{TNG0053_merger_history}
\end{figure*}

We use the time evolution of TNG0053 to constrain the history of each component of the metallicity distribution.  We consider the history of the gas-phase oxygen metallicity gradient, gas surface density, stellar surface density, and star formation rate surface density of TNG0053 from $z=5$ to $z=0$ (\autoref{fig:Fig_Gradient_sims_z}) and the stellar mass history of TNG0053, including any secondary merging galaxies with stellar mass greater than $\log(M_{*}/M_{\odot}) > 4.5$ (\autoref{TNG0053_merger_history}).  
This analysis reveals the gradient history:\\

\noindent{\bf Inner bar gradient ($R<7$~kpc):} In the model, the innermost region, within a 7 kpc radius, experienced an increase in gas surface density around 12 billion years ago, leading to heightened star formation from redshift $z=4$ to $z=0$. This star formation process enriched the inner region, resulting in a steep inner metallicity gradient.  Between $z=5$ to $z=3$ (12.5 to 11.5 Gyr ago), TNG0053 merged with one $10^{6.6}~M_{\odot}$ galaxy and two $\sim 10^{7.2}$ \Msun\ galaxies.  These secondary galaxies had a larger total metallicity than TNG0053 at the redshift of their merger. However, galaxy collisions can cause disruptions to the gravitational potential of a spiral galaxy, triggering large-scale flows of low metallicity gas from the outskirts of the spiral galaxy into its central regions, impacting metallicity gradients \cite{Kewley06b,Kewley10,Rupke10}. Previous models show that gas flows can dilute central metallicity gradients for brief periods of time ($\sim 200$~Myr) while triggering central star formation that enriches the gas over $\sim1$~Gyr in the absence of large-scale galactic winds  \cite{Torrey12}.  

Interestingly, the metallicity gradient of the inner region within TNG0053 flattened between $z=0.5$ to $z=0$ with a reduction in the central metallicity from \OH$\sim 9.5$ to \OH$\sim 9.1$ alongside a reduction of star formation activity.  During this period, TNG0053 merged with a secondary $\sim 10^9$ \Msun\ galaxy and two secondary $\sim 10^{8.8}$ \Msun\ galaxies between $z=0.2$ to $z=0$. These mergers likely triggered low metallicity gas inflows but star formation has not re-enriched the central gas.  This suppression of star formation and subsequent impact on the central metallicity can be caused by galactic-scale outflows driven by AGN \cite{Torrey12}. \\

\noindent{\bf Main disk gradient ($7 \geq R \leq 17$~kpc):} The model main disk gradient (7-17 kpc) began forming at redshift $z=5$ through gas inflow from a minor merger with a $10^{6.6}~M_{\odot}$ galaxy. The merging galaxy can be seen in \autoref{fig:Fig_Gradient_sims_z} at a distance of $\sim 30$~kpc from the nucleus of the primary galaxy.  The incoming galaxy entered the merger with gas that was significantly more enriched (\OH$\sim 6.4-7.2$) than the outer regions of the primary galaxy (\OH$\sim 5.8-6.0$ between 15-20 kpc).  Subsequent mergers with two galaxies at $z=3.5$ that have a combined mass of $10^{7.5}~M_{\odot}$ and associated star formation further enriched the main disk.  From $z\sim3$ to $z\sim1$, the main disk gradient continues enriching through star formation during periods without mergers.\\

\noindent{\bf Flat outer disk ($17<r<28$ kpc):}   The model outer disk beyond 17 kpc exhibits a flat metallicity gradient formed relatively recently (between $z=1$ to $z=0.5$) due to the the infall of metal-rich gas from a merger with a dwarf galaxy with stellar mass of $10^{9}$~\Msun. This outer disk region expanded from 21 kpc to approximately 30 kpc in radius between redshifts $z=1$ and $z=0.5$ due to the inflow of gas from this merger.

Numerous possible causes of flat outer gradients have been proposed, including the flat gas surface densities in the outskirts of galaxies \cite{Bresolin09}, a radial dependence of star-formation activity \cite{Phillipps91}, redistribution of gas from recent mergers \cite{Bresolin09,Werk11}, a lack of evolution in the extended galaxy disks \cite{Bresolin09}, and metal transport to the outer regions of galaxies through a galactic fountain \cite{Werk11}.  Our simulated model suggests that the outer region of the gas disk was built up recently, with the ionized gas and stellar disk increasing from 21~kpc to $\sim 30$~kpc in radius 5-8 billion years ago (from $z=1$ to $z=0.5$) through the infall of gas and subsequent star formation and chemical enrichment from the merger with a more metal rich $10^{9}$~\Msun\ galaxy.   

The infall of gas in TNG0053 has likely been facilitated by the presence of the spiral arm structure.  Recent FIRE-2 simulations suggest that the gas within spiral arms can move inwards along metal-rich arms and outwards along secondary metal poor arms, providing an efficient mechanism for gas flows to and from the central regions \cite{Orr23}.  The steepness of the gradient in TNG0053 may also be related to the spiral structure; recent observational work shows that the presence of spiral structure correlates with steeper gas phase metallicity gradients \cite{Wisz25}.  

\subsection{Discussion}

The possible accretion and metallicity gradient history of NGC~1365 has, in this work, been explored through a single, limited set of cosmological simulations of galaxy formation and evolution (Illustris TNG).  We find a best-match to the stellar mass and metallicity gradient of NGC~1365 using the highest resolution Illustris TNG50 simulations.  We use the history of the metallicity gradient, star formation density, gas density, and stellar mass density of the matched model to trace a possible star formation and accretion history of NGC 1365.
The current comparison is limited to a small set of theoretical galaxy matches found in the small volume (51.7 Mpc) covered by the highest resolution Illustris simulations. A larger volume high-resolution simulation would enable a larger set of galaxy parameters to be matched.  A larger model grid would allow matches to additional properties such as azimuthal metallicity variations, the presence or absence of a bar, stellar metallicities, and star formation history.   

Azimuthal metallicity variations were discovered in NGC~1365 by \citet{Ho17}.  They showed that the \HII regions downstream of the spiral arms in NGC~1365 have a shallower metallicity gradient than the arm \HII\ region \cite{Ho17}).  In Figure~\ref{fig:NGC_TNG0053_images} we show the spatial metallicity variations in NGC~1365 and TNG0053. Both NGC~1365 and TNG0053 have azimuthal variations where the spaxels downstream from the spiral arms have lower metallicities and shallower gradients than the spaxels within the spiral arms.  A larger set of simulations would allow azimuthal variations to be quantitatively matched and studied, in addition to the one-dimensional metallicity gradient.

In this work, the closest match to NGC~1365 does not have a bar and it is likely that differences exist in the formation of the inner gradient between TNG0053 and NGC~1365.  In particular, the absence of a bar suggests the absence of the processes that regulate bar formation.  In addition, the efficiency of gas inflows and outflows may differ in barred and unbarred galaxies.  The small number of galaxies that match NGC~1365 do not allow a statistically significant investigation or quantification of this effect.  We will investigate the theoretical impact of barred and unbarred galaxies on metallicity gradients within the larger TNG sample in future work.

Different feedback models may impact the derived theoretical metallicity gradients.  The metallicity gradient production in smooth feedback models is not sensitive to feedback model parameters.  However, our recent work suggests that smooth stellar feedback may not sufficiently mix the metal content in galaxies and that either stronger stellar feedback or additional subgrid turbulent metal diffusion models may be required to more accurately reproduce the observed metallicity gradients in nearby galaxies \cite{Garcia25}. In future work, we will apply our technique to a larger sample of TYPHOON galaxies and explore the metallicity gradient history using multiple different cosmological simulations, including simulations with different feedback models.

Finally, the analysis presented here suggests that the individual accretion history of spiral galaxies can be explored using the archaeological record embedded in their metallicity gradients through the combination of ultra-high resolution 3D spectroscopy and high resolution cosmological simulations.  Galaxies that are prime targets for this technique are face-on spiral galaxies with 3D spectroscopy with spatial resolution of 175 pc or higher.

\section{Methods}\label{Methods}

\subsection{NGC~1365 observations and data reduction}

NGC~1365 was observed as part of the TYPHOON survey on the 2.5m du Pont telescope at the Las Campanas Observatory.  The observations and data reduction of NGC~1365 are described in \cite{Ho17}.  Briefly, a custom long slit (18' x 1".65) was stepped across the galaxy using the Wide Field CCD (WFCCD) imaging spectrograph.  The entire optical disk of NGC~1365 was covered within the 25 arcminute field of view.  The reduced long-slit 2D data were tiled together to form a 3D data cube.  The resulting spectra cover 3650-8150\AA\ with spectral sampling of 1.5\AA\ and spatial sampling of 1.65". The spectral resolution is R$\sim 850$ at 7000\AA. The full-width half-maximum (FWHM) point spread function is 175~pc at the distance of NGC~1365 \cite{Ho17}.  We emphasize that the high resolution of TYPHOON is crucial for accurately determining metallicity gradients \cite{Poetrodjojo19}. 

\subsection{NGC~1365 data analysis and derived quantities}

The emission-lines in each spaxel were fit using LZIFU \cite{Ho16b}.  LZIFU uses PPXF \cite{Cappellari17} to fit and subtract the stellar continuum and MPFIT to fit Gaussian profiles to the emission-lines via the least-squares technique \cite{Markwardt09}, providing emission-line flux maps and associated error maps.  Emission-line fluxes were dereddened using a Milky Way extinction curve \cite{Fitzpatrick99}.  We select spaxels with $S/N>3$ for \Ha~$\lambda 6563$, \Hb~$\lambda 4861$, \SII~$\lambda \lambda 6717,31$ and \NII~$\lambda6563$.  We remove 126 spaxels dominated by non-star forming ionizing radiation using the \citet{Kewley06a} classification scheme on the Baldwin, Phillips \& Terlevich (BPT) diagram \citep{Baldwin81}, leaving 4546 spaxels.  A total of 117/126 (93\%) of the removed spaxels are within the 3.5~kpc radius of the AGN extended emission-line region.

For comparison, we also identify \HII\ regions by applying the HIIphot package \cite{Thilker14}.  We identify a total of 283 \HII\ regions with $S/N>3$ in \Ha~$\lambda 6563$, \Hb~$\lambda 4861$, \SII~$\lambda \lambda 6717,31$ and \NII~$\lambda6563$.  Each spaxel and HII region associated with an independent galactocentric radius is corrected for inclination under the assumption of a circular thin disk. 

We calculate the star formation rate (SFR) of NGC~1365 using the TYPHOON \Ha\ emission-line and the \citet{Kennicutt94} relationship between the \Ha\ luminosity and SFR.  We use the flat $\Lambda$-dominated cosmology as measured by the 7 year WMAP experiment \cite[$h = 0.7$, $\Omega_m = 0.3$]{Komatsu11}.  Using the 283 \HII\ regions to calculate the SFR, we obtain an SFR(\Ha) of 5.67~${\rm M_{\odot}/yr}$.  

We apply metallicity diagnostics to the spaxels and \HII\ regions in NGC~1365.  For this work, we use the Dopita et al. (2016) \NIISII\ method due to its insensitivity to excitation, ionization parameter, ISM pressure and reddening, as well as the fact that the \NIISII\ diagnostic depends linearly on metallicity over the full metallicity range from $8.0<\log(O/H)+12<9.2$ \cite{Dopita16}.  We do not apply metallicity diagnostics using the \ratioR23\ ratio because this ratio is insensitive to metallicities in the critical range of $8.4<\log(O/H)+12<8.6$ \cite{Kewley02}. Ho et al. (2017)  calculated the metallicity distribution for NGC~1365 using four different metallicity diagnostics \cite{Ho17}. They showed that the metallicity distribution is unaffected by choice of metallicity diagnostic \cite[see Figures 10 and 11 in][]{Ho17}.  

We calculate the errors in the metallicities using the bootstrap resampling method with 1000 iterations within the metallicity measurement uncertainty.  Uncertainties in the metallicity from propagating line flux errors include the systematic uncertainties in the metallicity calibration and reddening correction.  Variations in metallicity at fixed radius may arise from localized enrichment events, turbulence, inflows, or variations in star formation history, all of which contribute to the overall scatter.  Because our scatter does not contain extreme outliers, we choose to represent our full scatter by the standard deviation, rather than the interquartile range, which uses only two data points.

\subsection{NGC~1365 metallicity gradient fitting method}

The metallicity gradient and the precise locations of any break radii are simultaneously determined with a brute-force fitting algorithm.  To avoid any assumptions about the metallicity gradient profile, we fit a single gradient, a two-component gradient, and a three component gradient.  We choose the profile that minimizes the $\chi^2$ across the metallicity profile as a function of radius.  For a two-component and three-component gradient, the break radius or radii are allowed to vary over the entire NGC~1365 metallicity profile, and profiles of the inner and outer zones are separately fit using linear regression.  For three-component fits, any identified candidate break-radius pair divides the profile into three distinct zones: an inner zone, an intermediate zone, and an outer zone.  We require the fit to be continuous at break radii, meaning that the fit parameters of the intermediate zone are constrained by the parameters of the inner and outer zones.  These fits are repeated for all candidate break-radius pairs, and the $\chi^2$ over the entire profile is measured for each fit. We find that only a three-component fit minimizes the $\chi^2$ of the fit.  We select the break-radius pair and corresponding gradients that minimize $\chi^2$ while exhibiting at least 2$\sigma$ deviation in metallicity gradients between successive zones.  Fit-parameter uncertainties of the inner and outer zones are derived from the fit covariance matrix, and uncertainties of the intermediate zone are derived from error propagation.

\subsection{IllustrisTNG simulations and matching method}

To understand the cause of the three-zone metallicity gradient in NGC 1365, we use the Illustris TNG cosmological simulation suite \citep[][hereafter TNG]{Marinacci18, Naiman18, Nelson18, Pillepich18b, Springel18, Pillepich19, Nelson19a, Nelson19b}.  Critical to the results in this work, the unresolved, dense, star-forming ISM in TNG is modeled with the \cite{Springel03} equation of state.  In this model, star particles are formed when the gas density becomes sufficiently high ($n_{\rm H}\gtrsim0.13\,{\rm cm}^{-3}$), with stellar masses following a \citet{Chabrier03} initial mass function. The stellar lifetime models are adopted from \cite{Portinari98} and depend on the stellar mass and metallicity.   As stars move off the main sequence, both mass and metals are returned to the ISM where the metal composition of the returned ejecta is set to match yields from detailed models of supernovae explosions (SNe) and asymptotic giant branch (AGB) wind composition \citep{Nomoto97, Kobayashi06, Karakas10, Doherty14, Fishlock14}. TNG tracks the evolution of nine chemical elements: H, He, C, N, O, Ne, Mg, Si, Fe, and uses a tenth chemical variable to monitor the total of all remaining untracked elements.  Once in the gas phase, the metals are advected with the fluid flow — potentially moving between mesh cells — but without any additional sub grid model for metal diffusion. For a complete description of the TNG model, see \citet{Weinberger17} and \citet{Pillepich18b}. 

The TNG suite is comprised of cosmological volumes with different resolutions.  The naming of the volumes corresponds, roughly, to the length of the cosmological box: TNG50 (51.7 Mpc)$^3$, TNG100 (110.7 Mpc)$^3$, and TNG300 (302.6 Mpc)$^3$.  In this work we use the highest resolution run, the TNG50 simulation (TNG50-1; hereafter synonymous with TNG), which has $2\times2160^3$ resolution elements, a mass resolution of $\sim10^5~M_\odot$ for baryonic particles, and contains $\sim 20,000$ galaxies at z$\sim 0$.  The star formation cell sizes of 100~pc are well-matched to the NGC~1365 FWHM point spread function of 175~pc.  Within TNG50, there are 111 galaxies that are within 0.2~dex of the stellar mass of NGC~1365 ($\log({\rm M / M}_{\odot}) = 10.75 \pm 0.10$;\cite{QianhuiChen23}).  Of these, 35 TNG galaxies have star formation rates between $3-7  {\rm M_{\odot}/yr}$, similar to the SFR of NGC~1365 of 5.67~${\rm M_{\odot}/yr}$.  A total of 26/35 of these TNG galaxies have bars.  

We search for TNG star-forming galaxies at redshift $z = 0$ for a match to: (1) the NGC~1365 spaxel metallicity gradient (\autoref{Fig_Gradient}) and (2) the NGC~1365 stellar mass of $\log({\rm M}_{*}/$\Msun$) = 10.80$ \cite{Ponomareva18}. To quantify the degree to which a given $z = 0$ TNG galaxy (in the appropriate stellar mass range) is similar to NGC~1365, we measure the $\chi^2$ between the NGC~1365 median metallicity profile and the TNG galaxy median metallicity profile. TNG galaxy metallicity profiles are measured via the methods described in \cite{Hemler21} and \cite{Garcia23}.  Because there is a non-zero offset between the TNG galaxy metallicity and the observed galaxy metallicity \cite[see][]{Nelson18}, we allow the TNG metallicity profile normalization to vary freely (including single, double, and triple gradient fits) and we consider only the normalization that produces the lowest $\chi^2$ value between the TNG profile and the observed profile.  The original profile has a zero-radius normalization of approximately 9.25 dex.

We find that TNG subfind ID 526478 (at snapshot 99) -- hereafter TNG0053 -- is by far the closest match to NGC 1365 out of the TNG50 sample of 20,000 simulated galaxies.  We apply the same brute force method to determine the break radii and metallicity gradients in TNG0053 as applied to NGC~1365.  The break radii in TNG0053 are located at 6.6~kpc and 16.5~kpc. The gradients within the inner steep bar region, the central disk region, and the outer disk regions are $-0.040 \pm 0.002$~dex~kpc$^{-1}$, $-0.018 \pm 0.003$~dex~kpc$^{-1}$, and $-0.000 \pm 0.002$~dex~kpc$^{-1}$, respectively.  TNG0053 has a SFR of 6.59~${\rm M_{\odot}/yr}$, similar to the SFR of NGC~1365 of 5.67~${\rm M_{\odot}/yr}$.

While TNG0053 minimizes the $\chi^2$ value between the simulated metallicity profile and the observed profile of NGC 1365, we also identified and inspected two other profiles with small $\chi^2$ values: TNG subfind IDs 526029 (TNG 0052; reduced $\chi^2 = 0.164$; log ${\rm M}_{*}=10.54$; SFR$=5.81 {\rm M_{\odot}/yr}$) and 537236 (TNG 0070; reduced $\chi^2=0.149$; log ${\rm M}_{*}=10.70$; SFR$=7.12 {\rm M_{\odot}/yr}$). Both models fail to adequately fit the NGC~1365 data.  TNG0052 has an inverted, steep outer gradient, and is a flocculent system. TNG0070 is lacking a flat outer gradient, and neither galaxy has a bar \cite{Zana22}.  TNG0052 and TNG0070 have higher reduced chi-squared values (0.164 and 0.149) than TNG0053 (0.123) and, more importantly, neither can reproduce the flat outer metallicity gradient observed in NGC 1365.  That only TNG0053 fits all constraints reinforces the power of high-resolution chemical mapping as an archaeological tool.

We use the dynamical history of TNG0053 to understand the formation of its metallicity gradient.  In this analysis, the lookback time is calculated using TNG redshifts and the Planck 2015 cosmology of ${\rm H}_{0} = 67.74$ and $\Omega_{\rm M} = 0.3089$ \cite{Planck16}.

\section{Data Availability}
The data that support the findings of this study are available on request from KG. The first public data release of TYPHOON will be made available through Data Central in 2026 (see https://typhoon.datacentral.org.au/Data-Access). 

\section{Code Availability}
All theoretical models used for this work are publicly available from the IllustrisTNG website (https://www.tng-project.org/data/).

\section{Acknowledgments}
We thank the three anonymous referees for their insightful comments on this manuscript. This paper is based on spectrophotometric data cubes obtained with the du Pont 2.5 m Telescope at the Las Campanas Observatory, in Chile, as part of the TYPHOON Program. K.G. is supported by the Australian Research Council through the Discovery Early Career Researcher Award (DECRA) Fellowship (project number DE220100766, KG) funded by the Australian Government and the Australian Research Council Centre of Excellence for All Sky Astrophysics in 3 Dimensions (ASTRO~3D), through project number CE170100013 (LJK).  PT and AG acknowledge support from NSF-AST 2346977 (PT) and the NSF-Simons AI Institute for Cosmic Origins which is supported by the National Science Foundation under Cooperative Agreement 2421782 (PT) and the Simons Foundation award MPS-AI-00010515 (PT).  This research has made use of the NASA/IPAC Extragalactic Database (NED), operated by the Jet Propulsion Laboratory, California Institute of Technology, under contract with NASA.  This research made use of NASA's Astrophysics Data System Bibliographic Services.  This research made use of Astropy a community-developed core Python package for Astronomy (Astropy Collaboration et al. 2013, 2018).   The authors acknowledge Research Computing at the University of Virginia and Harvard University for providing computational resources and technical support that have contributed to the results reported within this publication.

\section{Author contribution statement}
LJK conceived of and led the research and figure development. LJK wrote the manuscript. KG contributed to the NGC~1365 data reduction and error analysis, figure development, and manuscript preparation. AG aided the analysis of the TNG galaxies and related figures.  PT contributed to the theoretical interpretation of the NGC~1365 data and cultivation of the TNG simulation data.  JR carried out the observations of NGC~1365, developed the data reduction software, and assisted in data reduction.  ZSH conducted a preliminary metallicity analysis of TYPHOON galaxies and identified the analogs of TYPHOON galaxies in the TNG50 sample (including TNG0053).   Q-HC contributed to the fitting process for the radial gradient of NGC~1365 and the 2D metallicity maps.  PZ estimated the AGN-corrected star formation rate of NGC 1365 using the TYPHOON data.   MS contributed to the observations and assisted with the data reduction of the TYPHOON data cubes for NGC~1365.  LH led the development of the Illustris simulations and contributed theoretical expertise and to the interpretation of the observational data.  BM is the lead investigator and manager of the TYPHOON survey, including NGC 1365.   All co-authors provided feedback on the manuscript text.

\section{Competing interests statement}
The authors declare no competing interests.

\newpage

\end{document}